\documentclass[floatfix,eqsecnum,aps,nofootinbib, amssymb]{revtex4}

\begin{document}
\title{Neutrino masses in quartification schemes}
\author{Alison Demaria}\email{a.demaria@physics.unimelb.edu.au} \affiliation{School
of Physics, Research Centre for High Energy Physics, The University of
Melbourne, Victoria 3010, Australia} \author{Catherine I. Low}\email{c.low@physics.unimelb.edu.au} \affiliation{School
of Physics, Research Centre for High Energy Physics, The University of
Melbourne, Victoria 3010, Australia} \author{Raymond
R. Volkas}\email{r.volkas@physics.unimelb.edu.au} \affiliation{School
of Physics, Research Centre for High Energy Physics, The University of
Melbourne, Victoria 3010, Australia}

\begin{abstract}
The idea of quark-lepton universality at high energies has recently been explored in unified theories 
based upon the quartification gauge group $SU(3)^4$. These schemes encompass a quark-lepton exchange symmetry 
that results upon the introduction of leptonic colour. It has been demonstrated that in models in which 
the quartification gauge symmetry is broken down to the standard model gauge group, gauge coupling constant unification 
can be achieved, and there is no unique scenario. The same is also true when the leptonic colour gauge group is only 
partially broken, leaving a remnant $SU(2)_{\ell}$ symmetry at the standard model level. 
Here we perform an analysis of the neutrino mass spectrum of such models. We show that these models do not naturally generate 
small Majorana neutrino masses, thus correcting an error in our earlier quartification paper, but with the addition of one singlet neutral fermion per family there is a realisation of see-saw suppressed masses 
for the neutrinos.
We also show that these schemes are consistent with proton decay.
\end{abstract}

\maketitle

\section{introduction}
Quark-lepton universality at high energies has been advocated as one of the possible extensions to the standard model (SM). 
Within this class of extensions, a quark-lepton discrete symmetry is imposed after endowing the leptons with 
new {\it leptonic colour} degrees of freedom described by the separate gauge symmetry $SU(3)_{\ell}$. 
This idea was first explored by Foot and Lew \cite{Foot} with quarks and leptons being indistinguishable 
above an energy scale that could be as low as a few TeV. Within this context, the quark-lepton discrete symmetry does not 
require the unification of the gauge coupling constants. It is interesting that quark-lepton quantum number unification can be divorced from coupling constant unification, however, it is desirable to achieve the latter.  To accommodate such unification, the gauge symmetry may be enhanced 
to the ``quartification'' group $SU(3)_q \otimes SU(3)_L \otimes SU(3)_{\ell} \otimes SU(3)_R$ \cite{Ray, BMW}, 
augmented by a discrete symmetry relating the $SU(3)$ factors. 

This gauge group was independently explored by Volkas and Joshi \cite{Ray} and Babu, Ma and Willenbrock \cite{BMW}. 
The former model achieved only the 
partial unification of the gauge coupling constants with two independent gauge coupling constants at the 
unification scale. The latter found a way to achieve full unification through enforcing additional 
discrete symmetries and introducing non-renormalisable terms. This model also 
achieved suppressed nonzero neutrino masses with the addition of a heavy singlet per fermionic family. 
Both of these models had a residual $SU(2)_{\ell}$ symmetry at the SM level, 
a result of the leptonic colour symmetry being only partially broken. The motivation 
for our recent paper  \cite{us} was to see if full coupling constant quartification could be achieved without the introduction of additional symmetries and higher-dimension operators.  The basic change from Ref.~\cite{BMW} was to admit the possibility of intermediate symmetry breaking scales.

We performed a systematic analysis of the 
renormalisation-group equations (RGEs) for all possible symmetry breaking routes 
from the quartification gauge group down to the standard model gauge group $G_{SM}$. 
We demonstrated that full unification could be achieved using a number of these routes without the need for the 
additional discrete symmetry or the higher-dimension operators of \cite{BMW}. This was true both for models that break the symmetry down to $G_{SM}$ \cite{lbroken1, lbroken2} or only to $G_{SM} \otimes SU(2)_{\ell}$ \cite{Foot,Ray,BMW}. 
In the case where there was no residual $SU(2)_{\ell}$ symmetry, there were four 
independent choices that gave rise to unification of the gauge couplings, all of which predicted interesting, though distinct, phenomenology. 

The purpose of the present paper is to examine the neutrino sector of quartification models in more detail, and in the process to correct an error in our previous paper  \cite{us}.

The mass analysis of the leptonic sector for the models where $SU(2)_{\ell}$ was broken indicated that there were ten neutral leptonic components (per family) which, it was claimed, 
gained Majorana masses 
\cite{us}. Nine of the resulting mass eigenvalues were of order of the GUT scale, 
and the tenth was of order $\frac{M_{EW}^2}{M_{GUT}}$ for a generic $M_{GUT}$ and a single symmetry breaking event. 
This was the mass scale that would result 
from a regular see-saw mechanism \cite{seesaw} and this particle also displayed the correct weak coupling with the electron, 
suggesting that it was indeed a neutrino. 
It was claimed in this case that for a general model, see-saw suppressed masses for the neutrinos arose naturally, however 
it was anticipated that the introduction of intermediate scales, which would separate the values of the $M_{GUT}$ entries, 
would increase the value of this small eigenvalue. 

A more precise analysis of this neutral lepton sector was performed which sought to confirm see-saw suppressed masses 
for the neutrinos in the individual symmetry breaking cascades that achieved unification of the gauge coupling constants. 
In the course of this analysis, we discovered that two entries were incorrectly omitted from the Majorana mass matrix in the model 
with no residual $SU(2)_{\ell}$ symmetry of Sec. IIIB of Ref.~\cite{us}. 
The inclusion of these terms alters the general conclusion of Ref.~\cite{us} with regards to the neutrino masses. The results of the RGE analysis, however, remain unaffected with unification still achievable and not unique.    

Here we present a detailed analysis of the mass spectrum of the leptonic sector of these quartification models.
For the general case where there is only one symmetry breaking event and no residual $SU(2)_{\ell}$ symmetry, the 
mass eigenvalues reveal that there are two light (though not ultralight) eigenvalues, 
disagreeing with our previous claim. 
Here we propose a simple extension to ensure that only one ultralight (i.e.\ see-saw suppressed) active Majorana neutrino per family is generated \cite{seesaw}. It turns 
out that the introduction of a singlet fermion $S$ with
bare mass $M_S$ per family works well. Although the addition of these singlets slightly complicates the models, 
they are one way by which to generate see-saw suppression \cite{singlet}. 
In this paper we show that this is all that is 
required to generate the desired neutrino mass spectrum in the unified schemes of \cite{us}, with the 
additional discrete symmetries and higher-dimensional operators of \cite{BMW} still absent. 

A summary of the neutral sector of the quartification model with the gauge symmetry fully broken down to the 
standard model gauge group
is given in Sec.~\ref{cha:quart}. This contains details of the Yukawa sector and lists the four schemes 
which allow for the unification of the gauge coupling constants. In Sec.~\ref{cha:su2broken} we outline the mass analysis 
of the leptonic sector with and without the singlet present, and extend the analysis to consider the details of the 
four unified models. This analysis is repeated in the following section for the case where there is a residual $SU(2)_{\ell}$ 
symmetry. In Sec.~\ref{cha:proton} we estimate the lifetime of the proton and 
we conclude in Sec.~\ref{cha:conclusion}.

\section{The Quartification Model}\label{cha:quart}

The quartification gauge group is 
\begin{equation}
\centering
G_4 = SU(3)_q \otimes SU(3)_L \otimes SU(3)_{\ell} \otimes SU(3)_R.
\label{eqn:quart}
\end{equation} 
A $Z_4$ symmetry which cyclicly permutes the gauge groups as per $q \rightarrow L \rightarrow 
\ell \rightarrow R \rightarrow q$ is imposed, ensuring a single gauge coupling constant. 
The multiplets which house the fermions are represented by
$3 \times 3 $ matrices:
\begin{eqnarray}\label{eqn:fermions}
\centering
q &\sim& (\mathbf{3,\overline{3},1,1})=
\left(\begin{array}{ccc} 
d & u & h \\ 
d & u & h \\  
d & u & h \end{array} \right),\qquad
q^c \sim  (\mathbf{\overline{3},1,1,3})=
\left(\begin{array}{ccc} 
d^c & d^c & d^c \\  
u^c & u^c & u^c \\
h^c & h^c & h^c \end{array} \right),
\\ \nonumber
\ell&\sim&(\mathbf{1,3,\overline{3},1})=
\left(\begin{array}{ccc}x_1&x_2&\nu\\
y_1&y_2&e\\
z_1&z_2&N \end{array} \right),\qquad
\ell^c \sim  (\mathbf{1,1,3,\overline{3}})=
\left(\begin{array}{ccc}x_1^c&y_1^c&z_1^c\\
x_2^c&y_2^c&z_2^c\\
\nu^c&e^c&N^c\end{array}\right),
\end{eqnarray} 
where $q( \ell )$ contain the left-handed quarks(leptons) and  $q^c ( \ell^c )$ the left-handed antiquarks (antileptons). 
Both electrically neutral and charged exotic particles are needed to fill the representations. Of particular note, the exotic fermions
$x_i, y_i$ were required to be at the TeV scale in Ref.~\cite{BMW} to facilitate gauge coupling 
constant unification.

The Higgs fields are contained in two different $\mathbf{36}$'s of $G_4$, and are labelled as per
\begin{eqnarray}
\centering
\Phi_a & \sim (\mathbf{1, \overline{3},1,3}), \qquad 
\Phi_b \sim (\mathbf{3,1, \overline{3},1}), \qquad 
\Phi_c \sim (\mathbf{1,3,1, \overline{3}}), \qquad 
\Phi_d \sim (\mathbf{\overline{3},1,3,1}), \nonumber \\
 \Phi_\ell & \sim (\mathbf{1, 3, \overline{3},1}), \qquad 
\Phi_{\ell^c} \sim (\mathbf{1, 1,3, \overline{3}}), \qquad 
\Phi_{q^c} \sim (\mathbf{\overline{3},1,1,3}), \qquad 
\Phi_{q} \sim (\mathbf{3, \overline{3},1,1}).
\label{eqn:higgstrans}
\end{eqnarray}
These fields are closed under the $Z_4$ symmetry and generate 
realistic fermion masses and mixings, with the coupling to the fermions described by the $Z_4$-invariant 
Lagrangian\footnote{The notation $\ell \ell^c \Phi_a$ means $\overline{\ell}_R \ell_L \Phi_a$, etc.}
\begin{eqnarray}\label{eqn:lag}
\centering
&\mathcal{L}_{Yuk} &= Y_1 Tr \left( \ell \ell^c \Phi_a + \ell^c q^c \Phi_b + q^c q \Phi_c + q \ell \Phi_d \right) 
+ Y_2 Tr \left( \ell \ell^c \Phi_c^{\dagger} + \ell^c q^c \Phi_d^{\dagger} + q^c q \Phi_a^{\dagger} 
+ q \ell \Phi_b^{\dagger} \right)  \nonumber\\
&+&\, Y_L \epsilon^{jkm} \epsilon^{npr}  \left( \ell^{jn} \ell^{kp} (\Phi_{\ell})^{mr}
+   (\ell^c)^{jn} (\ell^c)^{kp} (\Phi_{\ell^c})^{mr} 
+   (q^c)^{jn} (q^c)^{kp} (\Phi_{q^c})^{mr}
+   (q)^{jn} (q)^{kp} (\Phi_{q})^{mr} \right) + H.c.
\end{eqnarray}
The Higgs fields are sufficient to achieve the necessary symmetry breaking, with the pattern 
$G_4 \stackrel{V}{\rightarrow} SU(3)_q \otimes SU(2)_L \otimes U(1)_Y \stackrel{u}{\rightarrow} SU(3)_q \otimes U(1)_Q$ 
accomplished by the fields obtaining VEVs of the form
\begin{equation}\label{eqn:VEV1}
\centering
\langle \Phi_{\ell} \rangle = \left(\begin{array}{ccc}
u_{\ell_1}&0&u_{\ell_2}\\
0&u_{\ell_3}&0\\
V_{\ell_1}&0&V_{\ell_2}\end{array}\right), \qquad 
\langle \Phi_{\ell^c} \rangle = 
\left(\begin{array}{ccc}
V_{\ell^c_1}&0&V_{\ell^c_2}\\
0&V_{\ell^c_3}&0\\
V_{\ell^c_4}&0&V_{\ell^c_5}\end{array}\right), \qquad 
\langle \Phi_{a} \rangle \sim \langle \Phi_{c}^{\dagger} \rangle = \left(\begin{array}{ccc}
u_{a_1}&0&u_{a_2}\\
0&u_{a_3}&0\\
V_{a_1}&0&V_{a_2}\end{array}\right).
\end{equation}
The $u$'s are of electroweak order and the $V_i$'s determine the higher energy breaking. 
Since $\Phi_a$ and $\Phi_c^{\dagger}$ have the same quantum numbers and hence VEV pattern, they can be treated
together for the purposes of this paper.

The four independent symmetry breaking cascades which were shown to permit unification 
are  
\begin{eqnarray}
\centering
1. \;\; G_4 & \stackrel{v}{\rightarrow} & SU(3)_q \otimes SU(2)_L \otimes SU(2)_{\ell} \otimes SU(2)_R \otimes U(1) \nonumber \\
& \stackrel{x}{\rightarrow} & SU(3)_q \otimes SU(2)_L \otimes U(1)_{Y} \label{eq:cas1}\\
2. \;\; G_4 & \stackrel{v}{\rightarrow} & 
SU(3)_q \otimes SU(2)_L \otimes SU(3)_{\ell} \otimes SU(2)_R \otimes U(1) \nonumber \\
& \stackrel{w}{\rightarrow} & SU(3)_q \otimes SU(2)_L \otimes SU(2)_{\ell} \otimes SU(2)_R \otimes U(1) \nonumber \\
& \stackrel{x}{\rightarrow} & SU(3)_q \otimes SU(2)_L \otimes SU(2)_{\ell} \otimes U(1) \nonumber \\
& \stackrel{y}{\rightarrow} & SU(3)_q \otimes SU(2)_L \otimes U(1)_{Y} \label{eq:cas2}\\
3. \;\; G_4 & \stackrel{v}{\rightarrow} & 
SU(3)_q \otimes SU(2)_L \otimes SU(3)_{\ell} \otimes SU(2)_R \otimes U(1) \nonumber \\
& \stackrel{w}{\rightarrow} & SU(3)_q \otimes SU(2)_L \otimes SU(2)_{\ell} \otimes SU(2)_R \otimes U(1) \nonumber \\
& \stackrel{x}{\rightarrow} & SU(3)_q \otimes SU(2)_L \otimes SU(2)_R \otimes U(1) \nonumber \\
& \stackrel{y}{\rightarrow} & SU(3)_q \otimes SU(2)_L \otimes U(1)_{Y} \label{eq:cas3}\\
4. \;\; G_4 & \stackrel{v}{\rightarrow} & 
SU(3)_q \otimes SU(2)_L \otimes SU(3)_{\ell} \otimes SU(2)_R \otimes U(1) \nonumber \\
& \stackrel{w}{\rightarrow} & SU(3)_q \otimes SU(2)_L \otimes SU(3)_{\ell} \otimes U(1) \nonumber \\
& \stackrel{x}{\rightarrow} & SU(3)_q \otimes SU(2)_L \otimes SU(2)_{\ell} \otimes U(1) \nonumber \\
& \stackrel{y}{\rightarrow} & SU(3)_q \otimes SU(2)_L \otimes U(1)_{Y} \label{eq:cas4}.
\end{eqnarray}
There are actually eight independent symmetry breaking routes from the quartification gauge group down to the standard model 
gauge group. Constraints on the unification of the gauge coupling constants revealed that four of these converged into the single 
possibility of Eq.~\ref{eq:cas1}, one disallowed unification entirely, and the other three are as listed in 
Eqs.~\ref{eq:cas2}-\ref{eq:cas4} \cite{us}.

With the introduction of these intermediate steps we now have five energy scales $v \ge w \ge x \ge y > u$, the VEV   
entries $V_i$ of Eq.~\ref{eqn:VEV1} now assuming the order 
$y,x,w$ or $v$ depending on the route. 
These are given by
\begin{eqnarray}
\centering
1. & V_{\ell_2,\ell^c_5,a_2} \sim v,  \qquad \qquad \qquad
V_{\ell_1,\ell^c_{1} ,\ell^c_{2},\ell^c_{3},\ell^c_4,a_1} \sim x,  \\
2. & V_{a_2} \sim v, \qquad V_{\ell_2,\ell^c_5} \sim w, \qquad V_{\ell^c_4,a_1} \sim x, \qquad 
V_{\ell_1,\ell^c_{1,2,3}} \sim y, \\
3. & V_{a_2} \sim v, \qquad V_{\ell_2, \ell^c_5} \sim w, \qquad V_{\ell_1,\ell^c_2} \sim x, \qquad 
V_{\ell^c_{1,3,4},a_1} \sim y, \\
4. & V_{a_2} \sim v, \qquad V_{a_1} \sim w, \qquad V_{\ell_2,\ell^c_4,\ell^c_5} \sim x,\qquad
V_{\ell_1,\ell^c_{1,2,3}} \sim y.
\end{eqnarray}
As the $V_i$'s assume different energies across the four cascades, the mass spectrum resulting from the Yukawa couplings of 
Eq.~\ref{eqn:lag} differs for each. This variation gives rise to distinct energy bounds on the intermediate scales when considering 
the RGEs \cite{us}. The scales of all the breakings which allow unification are listed in Table \ref{tab:unif1}.  
The first cascade offered no flexibility, with the unification occuring at a fixed value of energy scales only. The other 
three cascades deliver unification within a range of energies, with the last stage of breaking able to occur at $10^6$ GeV. 
Note that 
cascades two and three have identical energy thresholds despite being independent.

For details of the RGE analysis, including the exact VEV structure of Eq.~\ref{eqn:VEV1} and the fermion and scalar mass spectrum 
for each cascade, we refer the reader to Appendix B of Ref.~\cite{us}.
\footnote{The cascades from Ref.~\cite{us} have been relabelled here for convenience. Thus,
 here cascade one corresponds to cascades one, two, seven and eight of Ref.~\cite{us}, two is four, 
three is five and four represents six.} 
It is sufficient for the purposes of this paper to be aware of 
the energy bounds on the intermediate and unification scales, and recognise that the internal structure of the scalar VEVs, and hence 
fermion mass matrices will be unique for each cascade.  
\begin{table}
\begin{center}
\begin{tabular}{|c|c|c|c|c|c|}
\hline
Cascade  & $y$ & $x$ & $w$ & $v$  \\
\hline
1  &   & $7.1 \times 10^2 GeV$ &  & $1.3 \times 10^{13} GeV$\\
\hline
2 and 3 & $y_{min} \sim 1 TeV$ & $1TeV$ & $6.2 \times 10^{12} GeV$ &$v_{max} \sim 1.1 \times 10^{13} GeV$  \\
  &    & $4.2 \times 10^7 GeV$ & $4.2 \times 10^7 GeV$ &$v_{max} \sim 3.8 \times 10^{11} GeV$ \\
& $y_{max} \sim 1.2 \times 10^6 GeV$ & $1.2 \times 10^6 GeV$  &$1.2 \times 10^6 GeV$ &
$1.4 \times 10^{11} GeV$ \\
\hline
4 & $y_{min} \sim 1 TeV $ & $8.8 \times 10^{3} GeV$ &$3.6 \times 10^{10} GeV$ &
$v_{min} \sim 3.6 \times 10^{10} GeV$ \\
  &    & $4.2 \times 10^7 GeV$ & $4.2 \times 10^7 GeV$ &$v_{max} \sim 3.8 \times 10^{11} GeV$\\
 & $y_{max} \sim 1.2 \times 10^{6} GeV$ & $1.2 \times 10^{6} GeV$ &$1.2 \times 10^{6} GeV$ &
$1.4 \times 10^{11} GeV$ \\
\hline
\end{tabular}
\end{center}
\caption{The range of energies for the symmetry breaking scales that will 
consistently give unification of the gauge coupling constants. 
Cascade one features the last 
stage of symmetry breaking necessarily occuring below a TeV. The other three choices allow 
for a range in the intermediate scales while still preserving unification. 
When $y_{max}$ is chosen for cascades two, three and four, they also become equivalent. }\label{tab:unif1}
\end{table}

\section{Neutrino masses when $SU(2)_{\ell}$ is broken}\label{cha:su2broken}
Without restrictions imposed on the Higgs fields, the allowed Yukawa interaction terms in the leptonic sector are
\begin{equation}\label{eqn:higgscouplings}
\centering 
Y_1  Tr [ \ell \, \ell^{c} \, \Phi_{a} ],\qquad
Y_L \epsilon^{jkm} \epsilon^{npr}\ell^{jn} \ell^{kp} (\Phi_{\ell})^{mr}, \qquad
Y_L \epsilon^{jkm} \epsilon^{npr}(\ell^c)^{jn} (\ell^c)^{kp} (\Phi_{\ell^c})^{mr},
\end{equation}
with the $Y_2$ term effectively subsumed into the $Y_1$ term.
The leptons with a charge of $+1$ are the $e^c$, $y_1^c$, $z_2$ and $x_2$
components of Eq.~\ref{eqn:fermions}. They mix and form Dirac mass terms 
with the charge $-1$ lepton components $e$, $y_1$, $z_2^c$ and $x_2^c$,
in the manner  
\begin{equation}\label{leptonmass2}
\left(\begin{array}{cccc} e& y_1 & z_2^c & x_2^c\end{array}\right)
\left(\begin{array}{cccc}
u_{a_3} & 0 & -u_{\ell_1} & V_{\ell_1} \\
0 & u_{a_3} & u_{\ell_2} & -V_{\ell_2} \\
-V_{\ell^c_1} & V_{\ell^c_4} & V_{a_2} & u_{a_2} \\
V_{\ell^c_2} & -V_{\ell^c_5} & V_{a_1} & u_{a_1}
\end{array}\right)
\left(\begin{array}{c}
e^c\\y_1^c\\z_2\\x_2
\end{array}\right) + H.c.
\end{equation}
where all the Yukawa coupling factors have been absorbed into the VEVs.
There are three 
Dirac mass eigenvalues (per family) of the intermediate scales $v,w,$ or $x$, and one eigenvalue (per family) 
of electroweak scale corresponding to the $e$, $\mu$ and $\tau$ masses. 

The leptonic components $N$, $N^c$, $\nu$, $\nu^c$, $x_1$, $x_1^c$, $y_2$, $y_2^c$, $z_1$ 
and $z_1^c$ are all neutral. These ten neutral leptons gain 
Majorana masses, as per 
\begin{equation}
\centering
(\chi^c)^T \, 
\left(\begin{array}{cccccccccc}
0  & V_{a_2} & 0  & V_{a_1}& u_{\ell_3} & 0 & u_{\ell_1} & 0 & 0 & 0\\
V_{a_2} & 0  & u_{a_2} & 0 & 0 & V_{\ell^c_3} & 0 & V_{\ell^c_1} & 0 & 0 \\
0 & u_{a_2} & 0 &u_{a_1} & 0 & 0 &-V_{\ell_1} & 0 & -u_{\ell_3} & 0 \\
V_{a_1} & 0 & u_{a_1} & 0 & 0 & 0 & 0& -V_{\ell^c_2} & 0 & -V_{\ell^c_3} \\
u_{\ell_3} & 0 & 0 & 0 & 0& u_{a_1}  &V_{\ell_2} & 0&0& u_{a_2} \\
0 & V_{\ell^c_3} & 0 & 0 &u_{a_1} & 0 & 0 & V_{\ell^c_5} & V_{a_1} & 0 \\
u_{\ell_1} & 0 & -V_{\ell_1} & 0 & V_{\ell_2} & 0 & 0 & u_{a_3} & -u_{\ell_2} & 0 \\
0 & V_{\ell^c_1} & 0 & -V_{\ell^c_2} & 0 & V_{\ell^c_5} & u_{a_3} & 0 & 0
& -V_{\ell^c_4} \\
0 & 0 & -u_{\ell_3} & 0 & 0 & V_{a_1} & -u_{\ell_2} & 0 & 0 & V_{a_2} \\
0 & 0 & 0 & -V_{\ell^c_3}& u_{a_2} & 0 & 0 &-V_{\ell^c_4}  &V_{a_2} & 0
\end{array}\!\right)\nonumber
\, \chi,
\label{eq:10x10}
\end{equation}
where $\chi = \left(\begin{array}{cccccccccc}\!N&\!N^c&\!\nu&\!\nu^c&\!x_1&\!x_1^c&\!y_2&\!y_2^c&\!z_1&\!z_1^c\end{array}\!\right)^T.$
Eq.~\ref{eq:10x10} is the generic form of this mass matrix for all four cascades. Given that the $V_i$'s will have different 
orders depending upon the exact route, the strength of the mixings between particles will also vary across the four 
individual cascades, changing the internal structure of the matrix and hence eigenvalues. 
Although there are five energy scales within this matrix, the unification requirements give independence to only three 
of these. In the following analysis, the higher two scales $w$ and $v$ 
are constrained to be non-trivial functions of $x$ and $y$. 
Unfortunately, diagonalisation is not analytically tractable, with analytic approximations creating too much error in the 
energy regions of interest, so the relationship between the eigenvalues and the energy scales was ascertained numerically. 

Casacde one, with its fixed choice of energy scales, offers no choice of generating a small neutrino mass, revealing 
two light mass eigenvalues of $\mathcal{O}(u)$. These eigenvalues, although of the same order, are not identical, 
so they do not indicate a Dirac neutrino. 

At first glance, the other three cascades seem more promising given that there is flexibility in the energy scales. 
However, the mass eigenvalues of cascades two and three adopt a similar pattern, with two light eigenvalues which are again not 
identical. These eigenvalues are of $\mathcal{O}(u)$ and they do not exhibit any significant fluctuation from this 
order within the energy bounds listed in Table ~\ref{tab:unif1}. 

Cascade four is the only option which returns light eigenvalues which depart from the electroweak order. Again we have two 
similar, light eigenvalues with the discrepancies between their values more evident for low $x,y$. At this end of 
the symmetry breaking spectrum, we have a mass eigenvalue of no less than $\mathcal{O}(100)$ MeV, which increases to $\mathcal{O}(u)$ 
as $x$ and $y$ approach their maximum values. Thus, one can not obtain a small enough Majorana mass for the neutrino here. 

Additionally, in all of the above cases, the electron couples to two of the lightest eigenvalues, so there are 
two ``neutrino-like'' particles per family. Thus, we are unable to  
return nonzero suppressed neutrino masses in these models, and there are unwanted partners for the lightest
eigenstates also.

For schemes which are not forthcoming of a natural see-saw mechanism for neutrinos, there are three 
approaches by which to induce the correct relations which feature predominantly in model-building. 
These are the employment of discrete symmetries \cite{discrete}, 
the addition of a neutral fermion per family which transforms as a singlet under the gauge group \cite{singlet} 
and the implementation of non-renormalisable higher dimensional operators \cite{hdimops}. 
These three were all invoked in the model of Babu {\it et al.} \cite{BMW}, although not all were introduced to resolve 
the issue of neutrino mass. The former was required to forbid some of the exotic leptons gaining GUT scale masses, a result 
which would be incompatible with the unification of the gauge coupling constants. In this case, a small Majorana neutrino 
mass was obtained by the addition of a fermion singlet which was given a GUT scale bare mass. 

Unfortunately the minimal quartification models fail to automatically yield
see-saw suppressed neutrinos.  As the next best alternative, we try to construct the simplest satisfactory
non-minimal class of models.
The logical choice to consider is the addition of a singlet fermion $S$ with Majorana mass $M_S$. 
This action does not impact the 
renormalisation-group equation analysis of Ref.~\cite{us}, with the unification constraints still as per Table \ref{tab:unif1}. We shall consider a range of values for the mass $M_S$.  Being a fermion mass, naturalness
does not automatically prejudice us to take its value to be at the highest scale in the theory.
In fact, a small mass term 
for this singlet has been extensively employed to generate suppressed neutrino masses \cite{lightS}, and this was  
specifically shown in quark-lepton symmetric models in Ref.~\cite{lbroken2}. In the latter model, the 
neutrino mass term (family structure suppressed) is of the form
\begin{equation}\label{eqn:doubleseesaw}
\centering
\left( \begin{array}{ccc} \overline{\nu_L^c} & \overline{\nu_R} & \overline{S^c} \end{array} \right)
\left( \begin{array}{ccc} 0 & u & 0 \\
u & 0 & v \\
0 & v & M_S \end{array} \right)
\left( \begin{array}{c} \nu_L \\ \nu_R^c \\ S \end{array}\right).
\end{equation}
If $M_S=0$ then there is a massless Weyl neutrino and Dirac partner. For small, non-zero $M_S$ this 
Weyl state transforms into a Majorana neutrino with a mass twice suppressed by the GUT scale,
being of the order $u (u/v )(M_S / v)$.
Motivated by this, we try adding a singlet to our more complicated scenario.

The singlet couples to the neutral leptons via the gauge invariant terms
\begin{equation}\label{eqn:singlet}
\centering
\lambda_S \left( Tr \left[ \ell \Phi_{\ell}^{\dagger} \right] \,\, + \, \,  Tr \left[ \ell^c \Phi_{\ell^c}^{\dagger} \right]
\right) S.
\end{equation}
The mass matrix for the neutral leptons is now simply Eq.~\ref{eq:10x10} with the addition of a column and row corrsponding to S,
\begin{equation}
\centering
(\chi'^c)^T \, \,
\left(\begin{array}{ccccccccccc}
0  & V_{a_2} & 0  & V_{a_1}& u_{\ell_3} & 0 & u_{\ell_1} & 0 & 0 & 0 & \lambda_s V_{\ell_2}\\
V_{a_2} & 0  & u_{a_2} & 0 & 0 & V_{\ell^c_3} & 0 & V_{\ell^c_1} & 0 & 0 & \lambda_s V_{\ell_5^c}\\
0 & u_{a_2} & 0 &u_{a_1} & 0 & 0 &-V_{\ell_1} & 0 & -u_{\ell_3} & 0 & \lambda_s u_{\ell_2}\\
V_{a_1} & 0 & u_{a_1} & 0 & 0 & 0 & 0& -V_{\ell^c_2} & 0 & -V_{\ell^c_3} & \lambda_s V_{\ell_4^c}\\
u_{\ell_3} & 0 & 0 & 0 & 0& u_{a_1}  &V_{\ell_2} & 0&0&0 & \lambda_s u_{\ell_1}\\
0 & V_{\ell^c_3} & 0 & 0 &u_{a_1} & 0 & 0 & V_{\ell^c_5} & 0 & 0 & \lambda_s V_{\ell_1^c}\\
u_{\ell_1} & 0 & -V_{\ell_1} & 0 & V_{\ell_2} & 0 & 0 & u_{a_3} & -u_{\ell_2} & 0 & \lambda_s u_{\ell_3}\\
0 & V_{\ell^c_1} & 0 & -V_{\ell^c_2} & 0 & V_{\ell^c_5} & u_{a_3} & 0 & 0
& -V_{\ell^c_4} & \lambda_s V_{\ell_3^c}\\
0 & 0 & -u_{\ell_3} & 0 & 0 & 0 & -u_{\ell_2} & 0 & 0 & V_{a_2} & \lambda_s V_{\ell_1} \\
0 & 0 & 0 & -V_{\ell^c_3}& 0 & 0 & 0 & -V_{\ell^c_4} &V_{a_2} & 0 & \lambda_s V_{\ell_2^c}\\
\lambda_s V_{\ell_2} & \lambda_s V_{\ell_5^c} &\lambda_s u_{\ell_2} & \lambda_s V_{\ell_4^c} & 
\lambda_s u_{\ell_1}&\lambda_s V_{\ell_1^c} &\lambda_s u_{\ell_3}& \lambda_s V_{\ell_3^c}&\lambda_s V_{\ell_1}& \lambda_s V_{\ell_2^c}& M_S 
\end{array}\!\right)\nonumber \, \,
\chi',
\label{eq:11x11}
\end{equation}
with $\chi' = \left(\begin{array}{c}\chi \\ S \end{array} \right)$.
The same methodology for the eigenvalue analysis is now applied to this new mass matrix, however, now our search space expands to allow 
$M_S \in [1 \, TeV,v]$ as motivated above, and an arbitrary choice of the coupling $\lambda_s$. 

Again, cascade one can be immediately ruled out, the presence of the singlet not altering the pattern of 
the eigenvalues from the case above. The same is not true for the other three unification schemes.
Consider the second cascade. With the unification constraints imposed, both $v$ and $w$ are dependent on the $u,y$, and $x$ scales, giving 
us three variables. 
For low values of $y$ and $x$, the inclusion of the singlet does not have much of an observable effect on the 
eigenvalues from the scenario above. However, for larger values of $x$, and a small bare mass, we have one light 
mass eigenvalue. This occurs for $x \gtrsim 10^6 \, GeV$ and $M_S < 10^6 GeV$, with the mass becoming 
lighter for larger $x$ and smaller $M_S$. 
The most favourable outcome is when there exists a larger hierarchy between $x$ and $y$. 
Within this domain, $w$ and $v$ decrease, with $w \rightarrow x $ and $v/x \rightarrow 10^4$ GeV. 
For the range of energies that return a sole light 
eigenvalue, this particle displays the correct weak coupling with the electron and thus can be identified as the 
neutrino. 
Subsequently, if we consider a light singlet, we can generate 
see-saw suppressed neutrino masses for this choice of unification schemes.

Cascade three, although similar to two in its unification characteristics, does not offer as attractive a mass spectrum. 
The mass of the singlet must be less than $10^6$ GeV always to return a single light eigenvalue. However, although light, its value does 
not become ultralight unless we consider the limit $y \sim y_{min}$ and $x \sim 10^7 GeV$ when this choice becomes 
equivalent to cascade two. It is interesting to note that both cascades two and three require $M_S \lesssim \mathcal{O}(x)$ 
in order to return a ultralight eigenvalue. This suggests a non-trivial relationship between these two scales. This is not altogether 
surprising given that we have considered $w$ and $v$ to be functions of $x$, and so $x$ is the order parameter governing 
the lightness of this Majorana mass term. 

Cascade four also generates an ultralight neutrino mass. 
As long as $M_S \lesssim M_{GUT}$ we have two lighter Majorana eigenvalues. For larger values of $x$ we have a greater hierarchy 
in these two lightest eigenvalues for all values of $y\neq x$. 
Within this region, $x/w \rightarrow 1$ and the hierarchy between $v$ and $w$ increases.  
The lightest eigenvalue becomes more favourably light as the mass of the singlet is lowered, with an ultralight eigenvalue obtained for 
$M_S \lesssim 10^6$ GeV. This one light mass eigenvalue displays the correct weak coupling with the electron. 
In general, this eigenvalue is 
lighter for all values of the breaking scales than the previous cascades, giving us a lot more freedom. In particular, 
the last stage of symmetry breaking can occur at higher energies while still allowing for small neutrino mass. Additionally, 
as an ultralight eigenvalue can be generated for larger values of $x$, this pushes the unification scale to $10^{11}$ GeV, 
which, although still low, is more favourable than the lower unification bound of $10^{10}$ GeV. 

In all the above cascades, the presence of the singlet only has a significant influence on the two lightest eigenvalues,  
the remaining spectrum of particle mass maintaining consistency with that of Ref.~\cite{us} and Table \ref{tab:unif1}. 
Without an analytic form of the eigenvalues, it is difficult to ascertain precisely how the presence of the 
singlet influences the behaviour of the lightest eigenvalue and in particular, why $M_S < M_{GUT}$ yields a nice spectrum. 
In particular, the location of the coupling of order $x$ appears to, at least numerically, drive 
the order of this light eigenvalue. This arises as we have treated the scales undemocratically. 
Changing $x$ alters the relative hierarchies of all scales, and thus 
provides  
an energy domain in which there is see-saw suppression.  
From this analysis, it appears that Eq.~\ref{eq:11x11} is a (rather complicated) generalisation of Eq.~\ref{eqn:doubleseesaw}. 

In particular instances, analytic approximations of this light eigenvalue were obtained. However, in the 
energy regions of interest, these approximations became too erroneous. Nevertheless, inclusion of the singlet generates a 
hierarchy between the two lightest neutral mass terms, with the heavier of the two pushed up to an energy 
$ x > m_{\nu^c} > y$. This however, has no impact on the RGE analysis of Ref.~\cite{us}. 
In the regions in which we have an ultralight eigenvalue for $m_{\nu}$, the second lightest term, corresponding to 
$m_{\nu^c}$, has a value that in general 
approaches, but never reaches order $x$, and thus these values are consistent with the previous calculations. 
Even if this mass were to exceed order $x$, the resultant effect on the RGEs would be negligible. 
This means also that any new physics due to the presence of this exotic singlet will be above a TeV. 

So adding a light singlet fermion to the quartification models of Ref.~\cite{us} gives rise to a small Majorana mass 
for the neutrinos for a subset of these schemes. The success of these scenarios 
depends non-trivially on the relative hierarchies between the 
intermediate scales. It is interesting to note that the introduction of intermediary symmetry breaking stages in 
quartification models yields both gauge coupling unification \cite{us}, and see-saw suppressed neutrino masses upon 
the addition of a fermion singlet.  
The presence of this singlet, while a complication, does preserve
many of the attractive features of quartification.
Importantly, the previous RGE analysis is unaffected, giving us a unified scheme based on the 
quartification gauge group that can deliver the SM and predicts interesting new phenomenology at $1$ TeV and above. 
As the addition of the singlet has not added any new phenomenology below a TeV here, the phenomenological account of Sec.V of 
Ref.~\cite{us} is still valid.

\section{Neutrino masses when $SU(2)_{\ell}$ is unbroken}

We now consider the case where a residual $SU(2)_{\ell}$ symmetry remains at the standard model level. 
The four independent routes for this symmetry breaking are  
\begin{eqnarray}
\centering
1. \: \: \: G_4 \: \: \: \stackrel{v}{\rightarrow} & SU(3)_q & \otimes SU(3)_L \otimes SU(2)_{\ell} \otimes SU(2)_R 
\otimes U(1) \\ \nonumber
\stackrel{w}{\rightarrow} & SU(3)_q & \otimes SU(2)_L \otimes SU(2)_{\ell} \otimes SU(2)_R 
\otimes U(1) \\ \nonumber
 \stackrel{x}{\rightarrow} & SU(3)_q & \otimes SU(2)_L \otimes SU(2)_{\ell}
\otimes U(1)_{Y} \label{eqn:cascade1a}\\ 
2. \: \: \: G_4 \: \: \: \stackrel{v}{\rightarrow} & SU(3)_q & \otimes SU(2)_L \otimes SU(2)_{\ell} \otimes SU(3)_R 
\otimes U(1) \\ \nonumber
\stackrel{w}{\rightarrow} & SU(3)_q & \otimes SU(2)_L \otimes SU(2)_{\ell} \otimes SU(2)_R 
\otimes U(1) \\ \nonumber
 \stackrel{x}{\rightarrow} & SU(3)_q & \otimes SU(2)_L \otimes SU(2)_{\ell}
\otimes U(1)_{Y}\\ 
3.\: \: \:  G_4 \: \: \: \stackrel{v}{\rightarrow} & SU(3)_q & \otimes SU(2)_L \otimes SU(3)_{\ell} \otimes SU(2)_R 
\otimes U(1) \\ \nonumber
\stackrel{w}{\rightarrow} & SU(3)_q & \otimes SU(2)_L \otimes SU(2)_{\ell} \otimes SU(2)_R 
\otimes U(1) \\ \nonumber
 \stackrel{x}{\rightarrow} & SU(3)_q & \otimes SU(2)_L \otimes SU(2)_{\ell}
\otimes U(1)_{Y}\\ 
4.\: \: \: G_4 \: \: \: \stackrel{v}{\rightarrow} & SU(3)_q & \otimes SU(2)_L \otimes SU(3)_{\ell} \otimes SU(2)_R 
\otimes U(1) \\ \nonumber
\stackrel{w}{\rightarrow} & SU(3)_q & \otimes SU(2)_L \otimes SU(3)_{\ell} 
\otimes U(1) \\ \nonumber
 \stackrel{x}{\rightarrow} & SU(3)_q & \otimes SU(2)_L \otimes SU(2)_{\ell}
\otimes U(1)_{Y}, \label{eqn:cascade4a}
\end{eqnarray}
all of which allow for unification of the gauge coupling constants. 
The VEV pattern which achieves this breaking is 
\begin{equation}\label{eqn:SUlVEV}
\langle \Phi_{\ell} \rangle = \left(\begin{array}{ccc}
0 & 0 & u_{\ell}\\
0 & 0 & 0\\
0 & 0 & V_{\ell}\end{array}\right)\qquad 
\langle \Phi_{\ell^c} \rangle = \left(\begin{array}{ccc}
0 & 0 & 0\\
0 & 0 & 0\\
V_{\ell^{c}_{1}} & 0 & V_{\ell^{c}_{2}}\end{array}\right)\qquad 
\langle \Phi_{a} \rangle = \langle \Phi_c^{\dagger} \rangle = 
\left(\begin{array}{ccc}
u_{a_1} & 0 & u_{a_2}\\
0 & u_{a_3} & 0\\
V_{a_1} & 0 & V_{a_2}\end{array}\right).
\end{equation}
The relationship between the $V$'s and the breaking scales for each cascade are listed in Table \ref{tab:unif2} along 
with the energy bounds of each symmetry breaking event which allows for the unification of the gauge couplings.  
For full details of the derivation of these scales, including the exact VEV structure and mass spectrum for each cascade see 
Sec. IVA and Appendix A of Ref.~\cite{us}. 
\begin{table}[h]
\begin{center}
\begin{tabular}{|c|c|c|c|c|}
\hline
Cascade & $V$'s &  $x$ & $w$ & $v$  \\
\hline
1  & $V_{\ell_2^c} \sim v, \, V_{a_2,\ell} \sim w, \,$  & $x_{min}\sim 1 TeV$ & $2.7 \times 10^{12} GeV$ & $ 1.2 \times 10^{17} GeV$  \\
  & $ V_{a_1,\ell_1^c} \sim x$ & $x_{max} \sim 6.4 \times 10^7 GeV$ & $7.5 \times 10^{13} GeV$ & $ 7.5 \times 10^{13} GeV$ \\
\hline
2   & $V_{\ell} \sim v, \, V_{a_2,\ell_2^c} \sim w, \,$  
& $x_{min} \sim 6.5 \times 10^5 GeV$ & $6.5 \times 10^{5} GeV$ & $3.9 \times 10^{19} GeV$ \\
   & $ V_{a_1,\ell_1^c} \sim x$ &  $x_{max} \sim 6.5 \times 10^{7} GeV$ & $7.4 \times 10^{13} GeV$ & $7.4 \times 10^{13} GeV$  \\
 \hline
3  & $V_{a_2} \sim v, \, V_{\ell,\ell_2^c} \sim w, \, $ 
& $x_{min} \sim 6.3 \times 10^7 GeV$ & $ 7.7 \times 10^{13} GeV$ & $ 7.7 \times 10^{13} GeV$ \\
  & $V_{a_1,\ell_1^c} \sim x$ & $x_{max} \sim 4.9 \times 10^{10} GeV$ & $ 4.9 \times 10^{10} GeV$ & $ 7 \times 10^{12} GeV$ \\
\hline
4  & $V_{a_2} \sim v, \, V_{a_1} \sim w, \, $
& $x_{min} \sim 6.2 \times 10^8 GeV$ & $1.7 \times 10^{12} GeV$ & $1.7 \times 10^{12} GeV $  \\
   & $V_{\ell,\ell_2^c} \sim x$& $x_{max} \sim 4.8 \times 10^{10} GeV$ & $4.8 \times 10^{10} GeV$ & $7 \times 10^{12} GeV$ \\
\hline
\end{tabular}
\end{center}
\caption{Range of energy scales of symmetry breaking that yield unification of the gauge coupling constants when 
the quartification gauge group is broken to $G_{SM} \otimes SU(2)_{\ell}$. 
There is only one scenerio which allows for a TeV level breaking scale. 
The others occur at higher energies, with much narrower ranges for the final stage of breaking.}
\label{tab:unif2}
\end{table}

Again there are a range of possible energies at which unification and symmetry breaking are compatible, with the unification 
condition constraining $v$ and $w$ to be functions of $x$ and $u$ only. Only one choice allows for a TeV level breaking, the other 
three routes having a much narrower range for the final stage of breaking. 

The $x_1, x_1^c, y_2, y_2^c, z_1$ and $z_1^c$ components are now electrically charged and decouple from 
the $N, N^c, \nu, \nu^c$ components. The mixing between the charged sector is given by 
\begin{equation}
\centering
\left( \begin{array}{ccc} x_1 & z_1 & y_2^c \end{array} \right) 
\left( \begin{array}{ccc} u_{a_1} & u_{a_2} & V_{\ell} \\
V_{a_1} & V_{a_2} & -u_{\ell} \\
V_{\ell^c_2} & - V_{\ell_1^c} & u_{a_3} \end{array} \right) \left(\begin{array}{c}x_1^c \\ z_1^c \\ y_2 \end{array}\right) \, \, + \, \, 
\left( \begin{array}{ccc} x_2 & z_2 & y_1^c \end{array} \right) 
\left( \begin{array}{ccc} u_{a_1} & u_{a_2} & -V_{\ell} \\
V_{a_1} & V_{a_2} & u_{\ell} \\
-V_{\ell^c_2} & V_{\ell_1^c} & u_{a_3} \end{array} \right) \left(\begin{array}{c}x_2^c \\ z_2^c \\ y_1 \end{array}\right)\, \, + \,\,
u_{a_3} \, e\,e^c.
\end{equation}
There is one electroweak Dirac term (per family) corresponding to the $e, \mu$ and $\tau$ masses, and the exotic charged leptons 
obtain Dirac masses of order $V$.

As was concluded in Ref. \cite{us}, this model does not admit a Majorana mass for the neutrinos. 
Adding the singlet, the Majorana mass matrix has the form 
\begin{equation}
\left(\begin{array}{ccccc} N& N^c & \nu& \nu^c &  S \!\end{array}\right)
\left(\begin{array}{ccccc}
0  & V_{a_2} & 0  & V_{a_1}& \lambda_s V_{\ell}\\
V_{a_2} & 0  & u_{a_2} & 0 & \lambda_s V_{\ell_2^c}\\
0 & u_{a_2} & 0 &u_{a_1} & \lambda_s u_{\ell} \\
V_{a_1} & 0 & u_{a_1} & 0 &\lambda_s V_{\ell^c_1} \\
\lambda_s V_{\ell} & \lambda_s V_{\ell_2^c} &\lambda_s u_{\ell} & \lambda_s V_{\ell_1^c} & M_S 
\end{array}\!\right)\nonumber 
\left(\begin{array}{c}N \\ N^c \\ \nu \\ \nu^c \\ S \end{array}\right).
\label{eq:11x11b}
\end{equation}

For a generic symmetry breaking, the light eigenvalue can be approximated to be
\begin{equation}
\centering
m_{\nu} \approx \frac{ (u_{a_1}\,V_{a_2}-u_{a_2}\,V_{a_1})\left[ M_S ( u_{a_1}\,V_{a_2}\,-u_{a_2}\,V_{a_1})+
2 \lambda_s^2 (-u_{\ell}\,V_{a_2}\, V_{\ell^c_1}\,+u_{a_2}\,V_{\ell}\,V_{\ell^c_1}\,+u_{\ell}\,V_{a_1}\,V_{\ell^c_2}\, -
u_{a_1}\,V_{\ell}\,V_{\ell^c_2})\right]}{\lambda_s^2 \, (V_{a_2}\,V_{\ell^c_1}\,-V_{a_1}\,V_{\ell^c_2})^2}.\label{eq:light}
\end{equation}
If there were only one symmetry breaking event to $G_{SM} \otimes SU(2)_{\ell}$, then $V_i \rightarrow v$ and the light 
eigenvalue approaches $u^2/v$ in the limit $\lambda_s \rightarrow 1$. However, we know from the unification conditions of 
Table \ref{tab:unif2} that our intermediate scales are different and these influence the value of the light eigenvalue. 
In fact, it appears to be that the hierarchy between the intermediate scales aids in generating the lightness of this mass term, 
and Eq.~\ref{eq:light} remains valid only for certain subsets of the individual cascades upon identification of the $V$'s. 

For the first cascade the singlet does not conspire to give favourable results, with two light eigenvalues of 
order the electroweak scale for all possible values allowed by unification.

Turning now to cascade two where the value of the last stage of symmetry breaking has a much more narrow range. 
The analytic approximation 
of the light eigenvalue is valid only for 
$x \rightarrow x_{max}$. In this limit, we obtain two light eigenvalues, both of electroweak order. If $M_S \ll v$ then one of these 
becomes lighter and the other heavier, however, we still can not generate an ultralight mass. 
Thus we are concerned with 
the lower bound of the last stage of symmetry breaking. Fixing $M_S$ at the GUT scale, there is one ultralight mass eigenvalue for 
$x \lesssim 10^6 \,$ GeV. This light eigenvalue appears to be independent of $M_S$, remaining at approximately the same order for all 
choices of the singlet mass and displays the correct electroweak coupling. In this limit however, we have two additional light 
eigenvalues of electroweak order. The values of these two eigenvalues are significantly influenced only by $x$, not $M_S$, 
and become larger 
than a TeV in the region in which we no longer have an ultralight eigenvalue. 
In this case, the presence of the singlet alters the spectrum of our particles, with there being a light exotic singlet in addition to the 
right-handed neutrino and the SM neutrino. This does not impact the results of the RGE analysis of Ref.~\cite{us} as these exotic singlets 
can be taken to be the $N^c$ and $\nu^c$ components which carry hypercharge $Y=0$. 
So one must choose whether the generation of see-saw suppressed masses in this scenario warrants the presence of exotic singlets with 
masses under a TeV.

Cascade three, conversely, exhibits see-saw suppressed neutrino masses for values of $x$ closer to the upper energy bound. For smaller 
values of $x$, the presence of the singlet and the variation of its mass, does not prove fruitful. The Majorana mass matrix has 
three eigenvalues of order $v$ and two electroweak order eigenvalues which in the limit $ w \rightarrow v $ are approximately 
$  \frac{uv(v-x)}{v^2}$. Again, considering $M_S \sim v$, increasing $x$ affects three of the 
five eigenvalues. An order $v$ mass term becomes slightly lighter, approaching order $w$, while the hierarchy between the 
two lightest eigenvalues becomes more pronounced. As $x \rightarrow x_{max}$, the second lightest tends towards $10^6$ GeV, and the 
lightest becomes of order 
$\sim \frac{u^2 \left[ M_S - 2 \lambda_s^2 w(x) \right]}{\lambda_s^2 w(x)^2 }$ for smaller $M_S$ as given by Eq.~\ref{eq:light}.  
Lowering the singlet mass lowers the value 
of $x$ at which the hierarchy between the two lightest Majorana terms becomes prominent, and lowers the third eigenvalue closer to $w$. 
At $x=x_{max}$ we can generate an ultralight eigenvalue, the particle displaying the correct coupling with the electron. 
The second lightest eigenvalue 
approaches order $x$ but never attains it, and the limiting value of the heavier eigenvalue 
is larger than order $w$ always. This preserves the 
spectrum of masses considered for the RGE analysis in Ref.~\cite{us}, and importantly, there are no exotic 
singlets below a TeV.   

The final cascade is the only choice which returns the two lightest Majorana masses at distinct scales 
for all values of Table \ref{tab:unif2} irrespective of the 
scale of the singlet mass. 
In the limit $x \rightarrow w$, Eq.~\ref{eq:light} reduces to 
$\frac{u^2 M_S (v-w)^2}{\lambda_s^2 w^4}$, this 
mass becoming lighter for larger $x,w$. An ultralight eigenvalue can be obtained in this high $x$ limit, with the variation of 
the singlet mass not affecting this order much. This particle has the correct coupling with the electron. The other eigenvalues are 
influenced by the presence of the singlet, however, as in cascade three, the ordering of the particle spectrum is consistent with the 
RGE analysis, and the linear combination corresponding to $\nu^c$ gains a mass larger than a TeV. 

So when there is a residual $SU(2)_{\ell}$ gauge symmetry we still have three scenarios which achieve unification of the gauge 
coupling constants and generate neutrino masses that would result from the regular see-saw mechanism. 
One of these choices is less desirable, having exotic singlets at EW order. 
Again this appears to be a generalisation of the case of Ref~\cite{lbroken2}.
Furthermore, the range of energies in which this results demands $M_{GUT} > 10^{12} GeV$, meaning that Higgs induced proton 
decay has a greater suppression.
Although this result 
necessitates the addition of a fermion singlet per family, this is still an improvement on the model of Ref. \cite{BMW} as 
unification is still possible without the need to invoke a further discrete symmetry and non-renormalisable operators.  
In particular, this analysis cements that unification in this case is not unique. Additionally, for the latter two cascades, all exotic 
singlets lie above a TeV, thus in these scenarios there is no new phenomenology at a TeV other than that previously commented \cite{us}.

\section{proton decay}\label{cha:proton}
We finish with a brief comment on the proton decay in those models that return 
light Majorana masses for the neutrinos. Although in quartification models 
baryon number is conserved by the gauge interactions, proton decay may still 
proceed, mediated by Higgs boson exchange.  In the absence of $S$, baryon number conservation
can be imposed on the model.  With $S$ present, this is no longer the case, so we need to
examine the finite proton lifetime induced by the $S$ Yukawa coupling terms. 
Proton decay occurs through the 
mode $p \rightarrow \pi^+ \nu$. 
In addition to the terms in Eq.~\ref{eqn:singlet}, 
the singlet must also couple to the quark multiplets as per 
\begin{equation}\label{eqn:singlet3}
\centering
\lambda_s \left( Tr[q \Phi_q^{\dagger}] + Tr[q^c \Phi_{q^c}^{\dagger}] \right) S
\end{equation}
in order to be closed under the $Z_4$ symmetry. 
However, while the Yukawa Lagrangian of Eq.~\ref{eqn:lag} assigns $\Phi_q$ and $\Phi_{q^c}$ 
the baryon numbers $-\frac{2}{3}$ and $\frac{2}{3}$, respectively, the simultaneous presence
of the interaction terms of Eq.~\ref{eqn:singlet3} leads to baryon number violation by one unit. Proton
decay is mediated by a tree-level graph involving a virtual coloured Higgs field.  The decay products
must be $\pi^+ \nu$ because the field $S$ must be involved.

The current experimental lower bound on the lifetime of this decay mode is $\sim 10^{32}$ years \cite{lifetime}, with the 
partial width for this channel approximately
\begin{equation}
\centering
\Gamma \sim \frac{1}{32 \pi} (\lambda_s Y_L sin \, \theta )^2 \,\frac{ m_p^5}{m_{\Phi}^4}.
\end{equation}
$m_p$ and $m_{\Phi}$ are the masses of the proton and coloured Higgs fields respectively, 
$Y_L$ is the Yukawa coupling defined in Eq.~\ref{eqn:lag} 
and $sin \theta$ describes the admixture of $S$ within the light neutrino eigenstate. 
It turns out that this tiny mixing angle provides the necessary suppression. 

Given that both $\lambda_s$ and $\theta$ are not independent of the Majorana neutrino masses, it is not possible to 
make a generic estimation of the proton decay rate. Rather one must look at the cascades above individually. 
When the leptonic colour symmetry is broken entirely, there were only two choices which returned realistic neutrino masses. 
Looking first at cascade two, choosing the singlet coupling strength $\lambda_s \sim 10^{-3}$,  
a ultralight Majorana mass term results when the unification scale is $\sim 10^{11}$ GeV.  
The admixture of the singlet within this light neutrino eigenstate is $\theta \sim  10^{-4}$, 
giving an estimation of the lifetime to be $\tau \sim \frac{1}{Y_L^2} 10^{28}$ years. (In general for this case, 
over the range of allowable scales that gave unification, the 
mixing angle is $\theta \gtrsim 10^{-4}$). If the Yukawa coupling strength is $Y_L \lesssim 10^{-2}$, then the 
decay lifetime is above experimental bounds. 

Cascade four yields a more favourable estimate. For the same singlet coupling strength, the neutrino 
eigenstate has mixing angle $\theta \sim 10^{-6}$, and the unification scale is again of order $10^{11}$ GeV. 
This gives the lifetime of this particular mode to be $ \tau \sim \frac{1}{Y_L^2} 10^{32}$ years.   

When the $SU(2)_{\ell}$ symmetry remains unbroken, the unification scales are of larger orders and so generate 
a greater suppression of the decay rate independent of the mixing angle supression than the previous case. 
Cascade four returns the lowest of the unification scales and thus returns the shortest estimation of the lifetime. 
For symmetry breaking scales of $v \sim 10^{12}\, , \, w \sim 10^{11} \, , \, x \sim 10^9$ GeV 
and $\lambda_s = 10^{-2}$, 
the mixing angle is of order $\theta \sim 10^{-7}$. This gives an estimate 
of the proton lifetime to be $\tau \sim \frac{1}{Y_L^2} \, \, 4 \times 10^{36}$ years. So unless the Yukawa coupling 
$Y_L > 10^2$, then this lifetime is well above current experimental bounds.

\section{conclusion}\label{cha:conclusion}
Models based upon a quark-lepton symmetry offer alternative unification schemes. Gauge coupling constant 
unification can result in several of these schemes which consider different symmetry breaking routes from the 
quartification gauge group down to the standard model gauge group \cite{us}.
These schemes considered a minimal quartification scenario in which there were no additional symmetries 
or higher dimension operators. 
This paper investigated the neutrino sector of these successful unified schemes. 
We corrected an error in Ref.~\cite{us}, and as a consequence our claim in that paper that
light Majorana masses for the neutrinos can be generated 
in these minimal schemes was shown to be false.  To achieve a satisfactory neutrino mass outcome,
non-minimal models must therefore be considered.  We chose to add a fermion singlet per family.   
The addition of these singlets provided the necessary dynamics to return see-saw suppressed Majorana masses for the neutrinos 
in a subset of the unification schemes. When there was no residual leptonic colour symmetry, there were four unique cascades 
consistent with the gauge coupling unification. Of these, two returned both favourable neutrino masses and 
realistic bounds on proton stability. In the models that had a remnant $SU(2)_{\ell}$ symmetry, of the four possible choices, 
there were three which displayed consistency with both neutrino masses and proton decay. 

\acknowledgments{This work was supported in part by the Commonwealth of Australia and in part by the
Australian Research Council.}


\begin{thebibliography}{99}
\bibitem{Foot}
R.~Foot and H.~Lew, Phys.\ Rev.\ D {\bf 41}, 3502 (1990);
Nuovo Cim.\ A {\bf 104}, 167 (1991);
R.~Foot, H.~Lew and R.~R.~Volkas, Phys.\ Rev.\ D {\bf 44}, 1531 (1991).

\bibitem{Ray}
G.~C.~Joshi and R.~R.~Volkas, Phys.\ Rev.\ D {\bf 45}, 1711 (1992).

\bibitem{BMW}
K.~S.~Babu, E.~Ma and S.~Willenbrock,
Phys.\ Rev.\ D {\bf 69}, 051301 (2004).

\bibitem{us}
A.~Demaria and C.~I.~Low and R.~R.~Volkas, Phys. \ Rev. \ D {\bf 72}, 075007 (2005).

\bibitem{lbroken1}
R.~R.~Volkas, Phys.\ Rev.\ D {\bf 50}, 4625 (1994);

\bibitem{lbroken2}
R.~Foot and R.~R.~Volkas, Phys.\ Lett.\ B {\bf 358}, 318 (1995).

\bibitem{seesaw}
P.~Minkowski, Phys.\ Lett.\ B {\bf 67}, 421 (1977);
S.~L.~Glashow, in {\it Quarks and Leptons, Proceedings of the Advanced Study Institute} (1979);
M.~Gell-Mann, P.~Ramond and R.~Slansky, in {\it Supergravity}, ed.\ P. van Nieuwhenhuizen and
D.~X.~Freedman (North-Holland, Amsterdam, 1979), p.315;
T.~Yanagida, in {\it Proceedings of the Workshop on Unified Theories and Baryon Number
in the Universe}, ed.\ O.~Sawada and A.~Sugamoto (KEK Report No.\ 79-18, Tsukuba, Japan, 1979);
see also R.~N.~Mohapatra and G.~Senjanovic, Phys.\ Rev.\ Lett.\ {\bf 44}, 912 (1980).

\bibitem{singlet}
R.~N.~Mohapatra and J.~W.~F.~Valle, Phys.\ Rev.\ D {\bf 34}, 1642 (1986); 
 A.~S.~Joshipura and A.~Y.~Smirnov, Phys.\ Lett.\ B {\bf 439}, 103 (1998).

\bibitem{discrete}  
G.~C.~Branco and C.~Q.~Geng, 
 Phys.\ Rev.\ Lett.\  {\bf 58}, 969 (1987);
 B.~A.~Campbell, J.~R.~Ellis, K.~Enqvist, M.~K.~Gaillard and D.~V.~Nanopoulos,  Int.\ J.\ Mod.\ Phys.\ A {\bf 2}, 831 (1987);
 D.~London, G.~Belanger and J.~N.~Ng,  Phys.\ Rev.\ D {\bf 34}, 2867 (1986);
A.~S.~Joshipura and U.~Sarkar,  Phys.\ Rev.\ Lett.\  {\bf 57}, 33 (1986);
M.~Frank, I.~Turan and M.~Sher,  Phys.\ Rev.\ D {\bf 71}, 113001 (2005).

\bibitem{hdimops}  
S.~Nandi and U.~Sarkar, Phys.\ Rev.\ Lett.\  {\bf 56}, 564 (1986);
M.~Cvetic and P.~Langacker,  Phys.\ Rev.\ D {\bf 46}, 2759 (1992).

\bibitem{lightS}
D.~Wyler and L.~Wolfenstein, Nucl.\ Phys.\ B {\bf 218}, 205 (1983);
 R.~N.~Mohapatra and J.~W.~F.~Valle, Phys.\ Lett.\ B {\bf 177}, 47 (1986);
 R.~N.~Mohapatra,  arXiv:hep-ph/0306016.

\bibitem{lifetime}
S.~ Eidelman et al. (Particle Data Group), Phys. \ Lett. \ B {\bf 592}, 1 (2004).

\end{thebibliography}
\end{document}